\documentclass[prl,twocolumn,showpacs,amsmath,amssymb,superscriptaddress]{revtex4-1}

\usepackage{graphicx}% Include figure files
\usepackage{dcolumn}% Align table columns on decimal point
\usepackage{bm}     % bold math
\usepackage{multirow}
\usepackage{booktabs}
\usepackage{amsmath}
\usepackage{braket}
\usepackage{appendix}
\usepackage{float}
\usepackage{xcolor}
\usepackage{siunitx}
\usepackage{amssymb}
\usepackage{array}
\usepackage{hyperref}
\newcommand{\bb}[1]{\mathbf{#1}}
\newcommand{\md}{\text{ mod }}

\begin{document}
	\title{Superconductors with anomalous Floquet higher-order topology}

	\author{DinhDuy Vu}
	\affiliation{Condensed Matter Theory Center and Joint Quantum Institute, Department of Physics, University of Maryland, College Park, Maryland 20742, USA}

	\author{Rui-Xing Zhang}
	\email{ruixing@utk.edu}
	\affiliation{Condensed Matter Theory Center and Joint Quantum Institute, Department of Physics, University of Maryland, College Park, Maryland 20742, USA}
	\affiliation{Department of Physics and Astronomy, University of Tennessee, Knoxville, Tennessee 37996, USA}
	\affiliation{Department of Materials Science and Engineering, University of Tennessee, Knoxville, Tennessee 37996, USA}
	
	\author{Zhi-Cheng Yang}
	\affiliation{Joint Quantum Institute, University of Maryland, College Park, MD 20742, USA}
	\affiliation{Joint Center for Quantum Information and Computer Science, University of Maryland, College Park, MD 	20742, USA}

	\author{S. Das Sarma}
	\affiliation{Condensed Matter Theory Center and Joint Quantum Institute, Department of Physics, University of Maryland, College Park, Maryland 20742, USA}
	
	\begin{abstract}
		We develop a general theory for two-dimensional (2D) anomalous Floquet higher-order topological superconductors (AFHOTSC), which are dynamical Majorana-carrying phases of matter with no static counterpart. Despite the triviality of its bulk Floquet bands, an AFHOTSC generically features the simultaneous presence of corner-localized Majorana modes at both zero and $\pi/T$ quasi-energies, a phenomenon beyond the scope of any static topological band theory. We show that the key to  AFHOTSC is its unavoidable singular behavior in the phase spectrum of the bulk time-evolution operator. By mapping such evolution-phase singularities to the stroboscopic boundary signatures, we classify 2D AFHOTSCs that are protected by a rotation group symmetry in symmetry class D. We further extract a higher-order topological index for unambiguously predicting the presence of Floquet corner Majorana modes, which we confirm numerically. Our theory serves as a milestone towards a dynamical topological theory for Floquet superconducting systems.     
	\end{abstract}
	
	\maketitle
	
	\textit{Introduction -} During the past decades, topological superconductors (TSCs) have become a subject of great attention in the condensed matter community for hosting zero-dimensional (0D) anyonic Majorana quasiparticles, a fundamental building block for topological quantum information processing~\cite{Nayak2008,Bomantara2020}. Such 0D Majorana modes have been proposed to emerge as either end-localized bound states for a 1D $p$-wave TSC~\cite{Kitaev2001,Lutchyn2010} or as vortex bound states for a 2D $p+ip$ chiral TSC~\cite{read2000paired,fu2008}.       
	Notably, the quest for realizing Majorana physics also extends to dynamical systems, especially to periodically-driven Floquet systems. Remarkably, driving a 1D superconducting Bogoliubov-de Gennes (BdG) system can enable an ``anomalous" Floquet TSC phase~\cite{Jiang2011}, featuring the simultaneous presence of Majorana end modes at both $0$ and $\pi/T$ quasienergies. While such exotic Majorana phenomena are impossible in any equilibrium superconductors, it does not arise from the topology of the stroboscopic Floquet bulk bands, but is a direct consequence of the nontrivial quantum dynamics. Therefore, systems with similar anomalous Floquet topological properties are also known to be ``inherently dynamical"~\cite{Rudner,Nathan2015,fruchart2016complex,Roy2017} and are dynamically distinct from any static systems.
	
	Similar to the static higher-order topological insulators/superconductors \cite{benalcazar2017quantized,schindler2018higher,khalaf2018higher,miert2018higher} and their Floquet counterparts \cite{Bomantara2019floquet,Plekhanov2019floquet,Gosh2021floquet,Liu2021floquet}, the idea of anomalous Floquet topology has been extended to a higher-order version~\cite{Rodriguez-Vega2019afloquet,peng2019floquet,peng2020floquet,huang2020floquet,Bomantara2020afloquet,Zhang2020b,Zhu2021}. By definition, a $D$-dimensional Floquet system hosts $d$-th order anomalous Floquet higher-order topology (AFHOT) if it features $(D-d)$-dimensional boundary modes simultaneously occurring at both $0$ and $\pi$ quasienergies in the Floquet spectrum. In particular, the stability of AFHOT with $d>1$, in our particular case the anomalous Floquet second-order topology with $d=2$, is guaranteed only when there is an additional protection from crystalline symmetries. While AFHOT phenomena have been comprehensively studied for non-BdG systems~\cite{Zhang2020b}, little is known about their superconducting counterparts. In particular, a systematic topological classification for BdG systems with AFHOT is still missing.
	
	In this work, we provide a general topological theory for 2D class D anomalous Floquet higher-order topological superconductors (AFHOTSC) protected by $n$-fold rotation symmetry $C_n$. Such AFHOTSCs generically feature corner-localized Majorana bound states at both $0$ and $\pi$ quasienergies, which cannot be eliminated without breaking $C_n$ or closing the bulk quasienergy gap. In particular, we find that all AFHOTSCs necessarily manifest $C_n-$protected singularities in the phase spectrum of their time-evolution operators. Motivated by this observation, we first classify all possible phase-band singularities with a set of newly defined topological charges. By relating these singularities to those of 1D anomalous Floquet TSCs under dimensional reduction, we manage to establish a higher-order bulk-boundary correspondence between the phase-band singularities and the corner Majorana modes. This allows us to construct a new higher-order bulk topological index that can uniquely capture the presence or absence of anomalous Floquet corner Majorana modes. A minimal lattice model of a $C_2$-protected AFHOTSC is provided to demonstrate our general theory. 
	
	\textit{Phase band singularities -} Inherently dynamical systems are characterized by their inability of being smoothly deformed to static systems. For a time-periodic Hamiltonian $H({\bf k}, t+T) = H({\bf k}, t)$, such an obstruction can be visualized by calculating the phase bands $\{\phi_n ({\bf k}, t)\}$ of the time-evolution operator $U({\bf k}, t) = {\cal T} e^{-i\int H(\bb{k},t')dt'}$, which is defined by $U({\bf k}, t)|\phi_n({\bf k}, t)\rangle = e^{i\phi_n({\bf k}, t)} |\phi_n({\bf k}, t)\rangle$. Here ${\cal T}$ is the time-ordering operator. By definition, the phase bands $\{\phi_n ({\bf k}, t)\}$ are periodic along the ``energy" direction and we should only focus on the principal zone $(-\pi,\pi]$. In a pioneering work~\cite{Nathan2015}, it was demonstrated that the aforementioned obstruction to static systems can be attributed to the presence of phase-band singularities around principal-zone boundary. This theory successfully explains the origin of first-order anomalous Floquet topological phenomena for general tenfold-way symmetry classes. Similarly, this idea of phase-band singularities can be generalized to inherently dynamical systems with AFHOT~\cite{Zhang2020b}. This is the starting point for our classification scheme for $C_n$-symmetric AFHOTSCs in 2D. 
	
	We define a return map operator $\tilde{U}(\bb{k},t) = U(\bb{k},t) U(\bb{k},T)^{-t/T}$ \cite{Rudner}, which is time-periodic with $\tilde{U}(\bb{k},t+T) = \tilde{U}(\bb{k},t)$. In particular, the phase-band singularities of $U({\bf k}, t)$ must also exist in the phase spectrum of $\tilde{U}(\bb{k},t)$, since $U$ and $\tilde{U}$ essentially contain equivalent time-evolution information. Throughout our discussion, the term ``phase bands" will be referred to those defined for $\tilde{U}$, unless otherwise specified. For a 2D driven system, the return map spectrum features 3D dispersions with an additional dimension of time. Therefore, the most natural choice for a 3D robust singularity in the phase band is a 3D Weyl node carrying an integer-valued topological monopole charge. Indeed, as shown in Ref.~\cite{Nathan2015}, the net Weyl monopole charge of phase-band singularities $\nu_1\in\mathbb{Z}$ is a topological invariant for a general 2D Floquet system, which exactly equals the number of anomalous chiral edge modes in the quasienergy spectrum. Besides, similar to the cases of static higher-order topological phases~\cite{schindler2018higher,miert2018higher,khalaf2018higher}, spatial lattice symmetry (i.e., $C_n$ rotation symmetry in our case) is crucial for protecting AFHOTSC phases, which, however, is missing in the general definition of Weyl monopole charge of 3D phase-band singularities. It is then necessary to generalize the definition of monopole charge to a symmetry-representation-dependent version. 
	
	For 2D $C_n$-symmetric Floquet superconductors, the possible Weyl-like phase-band singularities can either live on the $C_n$ rotational axes or away from these high-symmetry axes. For the on-axis singularities, they are well described by a Weyl Hamiltonian under a basis of $C_n$-eigenstates $|\Psi_J\rangle = (\psi_{J}, \psi_{J+1})^T$ with $C_n\psi_{J}=e^{iJ\theta_0}\psi_{J}$ and $\theta_0=2\pi/n$. Here the angular momentum $J$ can be either integer or half-integer, depending on whether the system is spinless or spinful. In this basis, around a singularity at $\mathbf{K}=(\mathbf{k}_0,t_0)$, the effective phase-band Hamiltonian $\tilde{h}(\bb{k},t)=i\log \tilde{U}(\bb{k},t)$ is
	\begin{equation}\label{Hamiltonian}
		\tilde{h}(\bb{k}_0+\bb{k},t_0+t) = \pi\mathbb{I}_2 + v_t t\sigma_3 + v_k(k_x\sigma_1-k_y\sigma_2).
	\end{equation}
	Following Ref.~\cite{Zhang2020b}, we now define a new $J$-dependent monopole charge $q^\mathbf{K}_J=-\text{sgn}(v_t)$, which is essentially the Weyl monopole charge decorated with an additional representation-dependent label $J$, directly following our basis choice of the phase bands. 
	
	On the other hand, the Weyl singularities, if living off the rotational axes, must appear in $C_n$-related groups. Such a group of Weyl points can always be moved to a rotational axis in a symmetric way \cite{Supplement}. Without loss of generality, we hereafter assume in our later discussion that all singularities are located on the rotational-invariant axis. 
	
\begin{figure}
		\includegraphics[width=0.45\textwidth]{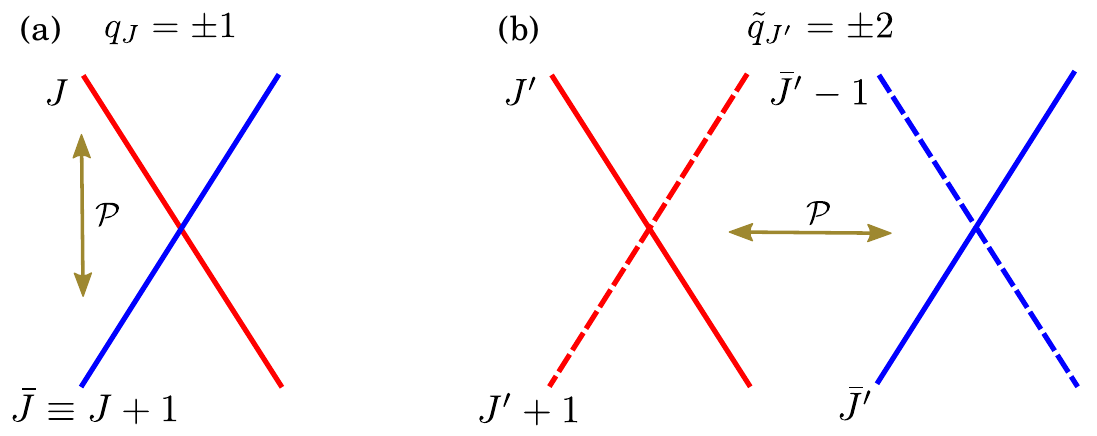}
		\caption{(a) A phase-band Weyl point that is invariant under particle-hole symmetry ${\cal P}$. (b) A PH-invariant Weyl pair where $\mathcal{P}$ maps one Weyl point to another.}\label{fig:Weylpoint}
	\end{figure}

	The particle-hole (PH) symmetry $\mathcal{P}$ for Floquet superconductors places an important constraint on the mathematical structure of these singularities. In particular, ${\cal P}$ acts on the phase-band Hamiltonian as ${\cal P} \tilde{h} ({\bf k}, t) {\cal P}^{-1} = -\tilde{h}(-{\bf k}, t)$, which flips only the sign of ${\bf k}$, but not that of $t$. Meanwhile, the $n$-fold rotation symmetry of a superconductor generally takes the form  $C_n=\text{diag}(R_n, e^{i\gamma\theta_0}R_n^*)$,
	where $R_n$ and $e^{i\gamma\theta_0}R_n^*$ denote the rotation matrix for the electron and hole parts, and $\gamma\in\{0,\dots,n-1\}$ signals a shift of angular momentum between electrons and holes. Importantly, for a \textit{single} Weyl node shown in Eq.~\eqref{Hamiltonian}, to preserve PH symmetry the two states $\psi_{J}$ and $\psi_{J+1}$ forming this Weyl node must be PH partners of one another [see Fig.~\ref{fig:Weylpoint} (a)], leading to the requirement
	\begin{equation}\label{eq:self_PH}
		J+1 \equiv \bar J \equiv -J+\gamma \md n,
	\end{equation}
	where we use the notation $\bar{J}$ as the index for the hole partner of the band $\psi_J$.
	Equation~\eqref{eq:self_PH} has two general solutions $J_1=(\gamma-1)/2$ and $J_2=(\gamma-1)/2+n/2$. As will be shown later, a rotationally symmetric system with $(C_n)^n=1(-1)$ can host AFHOT only when both $J_1$ and $J_2$ are simultaneously integers (half-integers). Since $J_2=J_1+n/2$, we immediately conclude that $C_3$ symmetry cannot protect AFHOTSC for failing to fulfill the above condition. 
	
	\begin{table}
		\centering
		\begin{tabular}{>{\centering} m{0.07\textwidth}  >{\centering}   m{0.07\textwidth}  >{\centering}  m{0.2\textwidth} >{\centering\arraybackslash} m{0.1\textwidth}}
			\toprule \\[-1em]
			$C_n$ & $\gamma$ &$(C_n)^n=-1$ & $(C_n)^n=1$\\
			\hline \\[-1em]
			\multirow{2}{*}{2} & 0 & $\mathbb{Z}\times\mathbb{Z}_2$ & $\mathbb{Z}$\\
			& 1 & $\mathbb{Z}$ & $\mathbb{Z}\times\mathbb{Z}_2$\\
			\hline \\[-1em]
			\multirow{3}{*}{3} & 0 &  $\mathbb{Z}$  &  $\mathbb{Z}$\\
			& 1 &  $\mathbb{Z}$  &  $\mathbb{Z}$ \\
			& 2 &  $\mathbb{Z}$  &  $\mathbb{Z}$\\
			\hline \\[-1em]
			\multirow{4}{*}{4}
			& 0 & $\mathbb{Z}\times\mathbb{Z}_2$ &  $\mathbb{Z}$\\
			& 1 &  $\mathbb{Z}$  & $\mathbb{Z}\times\mathbb{Z}_2$\\
			& 2 & $\mathbb{Z}\times\mathbb{Z}_2$ &  $\mathbb{Z}$\\
			& 3 &  $\mathbb{Z}$  & $\mathbb{Z}\times\mathbb{Z}_2$\\
			\hline \\[-1em]
			\multirow{6}{*}{6} 
			& 0 & $\mathbb{Z}\times\mathbb{Z}_2$ &  $\mathbb{Z}$\\
			& 1 &  $\mathbb{Z}$  & $\mathbb{Z}\times\mathbb{Z}_2$\\
			& 2 & $\mathbb{Z}\times\mathbb{Z}_2$ &  $\mathbb{Z}$ \\
			& 3 &  $\mathbb{Z}$  & $\mathbb{Z}\times\mathbb{Z}_2$ \\
			& 4 & $\mathbb{Z}\times\mathbb{Z}_2$ &  $\mathbb{Z}$ \\
			& 5 &  $\mathbb{Z}$  & $\mathbb{Z}\times\mathbb{Z}_2$ \\
			\bottomrule
		\end{tabular}
		\caption{Classification of 2D anomalous Floquet topological superconductors, where the first $\mathbb{Z}$ and the second $\mathbb{Z}_2$ indices indicate first-order and higher-order topology, respectively.\label{Tab:classification}} 
	\end{table}
	
	When $\psi_{J'}$ and $\psi_{J'+1}$ are not related by PH symmetry, Eq.~\eqref{eq:self_PH} does not apply and a phase-band Weyl node is necessarily accompanied by its PH partner, forming a PH-related Weyl pair [see Fig.~\ref{fig:Weylpoint} (b)]. The basis of such a Weyl pair $(\psi_{J'},\psi_{J'+1},\psi_{\bar{J'}-1},\psi_{\bar{J'}})^T$ are uniquely labeled by the following constraints: for $(C_n)^n=1$ $J'\in \{\lceil\gamma/2\rceil,\dots,\lfloor (\gamma+n)/2\rfloor-1\}$, thus $J'\in\varnothing$ if $n=2,$ 4 and $\gamma$ is even; for $(C_n)^n=-1$ systems $J'\in \{\lceil\gamma/2\rceil+1/2,\dots,\lfloor (\gamma+n)/2\rfloor-1/2\}$, thus $J'\in\varnothing$ if $n=2$, 4 and $\gamma$ is odd, with $\lfloor x \rfloor$ and $\lceil x \rceil$ being the floor and ceiling functions.
	In this basis, the PH-related Weyl pair at location $\mathbf{K}_0$ can be effectively described by a $4\times 4$ Hamiltonian $\tilde h(\bb{k}_0+\bb{k},t_0+t) = \pi\mathbb{I}_4+v_t t\Gamma_{45} +v_k(k_x\Gamma_4-k_y\Gamma_5)$. Here we have defined $\Gamma_i = \sigma_i\otimes\sigma_3$, $\Gamma_4=\sigma_0\otimes\sigma_1$, $\Gamma_5=\sigma_0\otimes\sigma_2$ and $\Gamma_{ij}=[\Gamma_i,\Gamma_j]/(2i)$. Notably, these two PH-related Weyl points must carry the same $J$-dependent monopole charge, and we thus label the net charge of this Weyl-pair as $\tilde{q}^\mathbf{K}_{J'} =-2\text{sgn}(v_t)$. Here we have applied a new notation $\tilde{q}_J$ to distinguish from the charge $q_J$ of a single PH-invariant Weyl singularity. Such PH-related Weyl pairs, together with the PH-invariant Weyl nodes, serve as the fundamental building blocks for all possible phase-band singularities that are compatible with both ${\cal P}$ and $C_n$ in 2D.
	
	\textit{$C_n$ topological charge -} Since a non-zero net Weyl charge indicates first-order topology with chiral edge modes, we hereafter restrict ourselves to cases with zero net Weyl charge with possible manifestation of second-order topological corner modes. It is crucial to note that two singularities can annihilate each other only when (i) their $J$-dependent monopole charges carry the same $J$ index; (ii) they have opposite $q_J$ or $\tilde{q}_J$. This annihilation constraint applies to both Weyl points and Weyl pairs. To see this, we consider merging two singularities that carry charges $q_{J_a}$ and $q_{J_b}$, respectively, with $q_{J_a} + q_{J_b} =0$. The phase-band Hamiltonian of the two singularities is described by 
	\begin{equation}\label{eq:2Weylpoints}
		\begin{split}
			\tilde{h}(\bb{k}_0+\bb{k},t_0+t) = \pi\mathbb{I}_4 + v_k(k_x\Gamma_4-k_y\Gamma_5)+ v_t t \Gamma_{3},
		\end{split}
	\end{equation} 
	under the basis $(\psi_{J_a},\psi_{J_a+1},\psi_{J_b},\psi_{J_b+1})^T$.
	Since $C_n\propto\sigma_0\otimes\text{diag}(1,e^{i\theta_0})$, when $J_a=J_b$, there exists a PH-symmetry-preserving mass term that is proportional to $\Gamma_1$, allowing the combined singularities to be gapped out. On the other hand, if $J_a\neq J_b$, such a mass term is not allowed by $C_n$-symmetry, and the pair of singularities is thus protected.
	
	One notable implication of our discussions so far is that a PH-related Weyl pair can never annihilate two Weyl points that are individually PH-invariant, since the allowed angular momenta for a PH-related Weyl pair and a PH-invariant Weyl point do not overlap.
	%$\{J'\}\neq J_{1,2}$ by definition. 
	This inspires us to define new independent $C_n$ topological charges to take into account all on-axis singularities,
	\begin{equation}\label{eq:invariant}
		Q_{J_{1,2}} = \sum_{\bb{K}} q^\bb{K}_{J_{1,2}},\quad \tilde{Q}_{J'} = \sum_{\bb{K}} \tilde{q}^\bb{K}_{J'},
	\end{equation}
	where we have summed over all possible rotational axes. The $\mathbb{Z}$-index of the anomalous first-order Floquet topology is $\nu_1 = Q_{J_1}+Q_{J_2}+\sum_{J'}\tilde Q_{J'}$, assuming there is no off-axis singularity.
	
	\begin{figure}
		\includegraphics[width=0.46\textwidth]{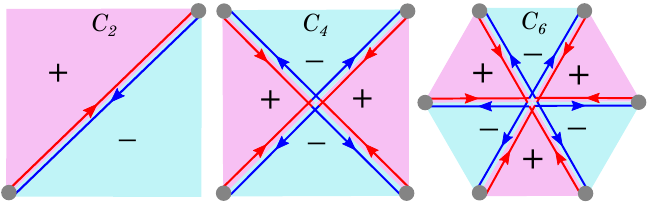}
		\caption{Coupling a globally $C_n-$symmetric mass term for a pair of oppositely-charged Weyl singularities leads to domain wall modes denoted by the arrows. Such a configuration of domain-wall modes directly indicates the presence of corner-localized $0$ and $\pi$ Majorana modes, whose spatial locations are denoted by dots.}\label{fig:dim_reduce}
	\end{figure}
	
	\textit{Bulk-boundary correspondence -} While the topological charges in Eq.~(\ref{eq:invariant}) provides a bulk classification of phase-band singularities, they do not directly lead to a classification of AFHOTSCs. One still needs to further translate these bulk topological charges into the corresponding boundary signature of the system.	To establish this precise higher-order bulk-boundary correspondence, we now exploit the idea of phase-band dimensional reduction (PBDR) that was first proposed in Ref.~\cite{Zhang2020b}. The essence of PBDR is to bridge between a higher-dimensional Floquet crystalline topological system and its lower-dimensional topological building blocks by introducing symmetry-allowed deformations to the phase-band singularities.

	To formalize this idea, we now consider the PBDR procedure for a general 2D $C_n$-symmetric Floquet superconductors with a pair of oppositely-charged phase-band singularities. This is the necessary setup for enabling AFHOT, and the zero-net monopole charge (i.e. $\nu_1=0$) guarantees the absence of chiral edge mode and hence first-order topology. Such a pair of singularities is effectively described by Eq.~\eqref{eq:2Weylpoints} with $J_a\neq J_b$. While we conclude that a homogeneous mass term $\propto\Gamma_1$ is $C_n$-forbidden, we now consider a time-independent but spatially varying mass term $m(r,\theta) \Gamma_1$ with
	\begin{equation}\label{eq:massterm}
		m(r,\theta)=m_0\Gamma_1\begin{cases}
			0  &\text{ if } \theta/\theta_0 \in \mathbb{Z} \\
			(-1)^{\lfloor \theta/\theta_0 \rfloor} &\text{ otherwise }
		\end{cases}.
	\end{equation}
	Here $r$ and $\theta$ denote the radial and angular directions in the polar coordinate, respectively. Remarkably, $m(r,\theta)$ indeed preserves the rotation symmetry as $C_nm(r,\theta)C_n^{-1}=-m(r,\theta)=m(r,\theta+\theta_0)$. As a result, the singularities are eliminated everywhere except along the $n$-fold domain walls of $m(r, \theta)$ at $\theta=l\theta_0$ for $l=0,1,...,n-1$, where the mass term flips its sign as shown in Fig.~\ref{fig:dim_reduce}. We show in \cite{Supplement} that such domain walls remain gapless against PH-invariant perturbations. Notably, these domain wall modes are precisely the phase-band singularities of the (1+1)D chains as a result of the dimensional reduction, which exactly correspond to 1D anomalous Floquet TSCs with end-localized Majorana modes at both $0$ and $\pi$ quasienergies~\cite{Nathan2015}. This directly implies that the original 2D $C_n$-symmetric Floquet superconductor must carry 0D Majorana modes localized at $C_n$-related corners, corresponding to $n$ pairs of Majorana zero and $\pi$ modes on the boundary (see Fig.~\ref{fig:dim_reduce}). 
	
On the other hand, neither (i) two Weyl pairs nor (ii) two Weyl points plus one Weyl pair with a zero net charge support stable gapless domain walls, which we show in Supplemental Materials ~\cite{Supplement}. The key distinction is that the singularities in this case are formed by eight phase bands, as opposed to four in the previous case. Therefore, only pairs of PH-invariant Weyl nodes may lead to robust corner Majorana modes, and hence higher-order topology. This immediately implies that $C_3$ symmetry cannot protect AFHOTSC, since Eq.~\ref{eq:self_PH} only has one solution for either spinful or spinless systems when $n = 3$, and thus there cannot be a pair of stable PH-invariant Weyl nodes with different $J$’s.

The generic case corresponds to combinations of the above-mentioned possibilities, where the singularity consists of a total number of $d = 8n$ or $d = 8n + 4$ phase bands. We thus conclude that AFHOT is present \textit{if and only if} $d/4$ is odd. In addition, we also require a vanishing net charge to exclude the possibility of first-order topology, with $Q_{J_1}+Q_{J_2}+\tilde{Q} = 0$ and $\tilde{Q}=\sum_{J'}\tilde{Q}_{J'}$ being an even number. We then find that    
	\begin{equation}
		\begin{split}
			d & = 2(|Q_{J_1}|+|Q_{J_2}|)+4|\tilde{Q}/2| \\
			& = 2(|Q_{J_1}|+|Q_{J_2}|+|Q_{J_1}+Q_{J_2}|).
		\end{split}
	\end{equation}
	Recall that the two values $J_{1,2}$ only exist for even $n$. Therefore, the condition for an odd $d/4$ can be simplified to a ${\mathbb Z}_2$ higher-order topological index
	\begin{equation}\label{eq:nu2}
		\nu_{2}=\frac{|Q_{J_1}|+|Q_{J_2}|-|Q_{J_1}+Q_{J_2}|}{2} \text{ mod }2, 
	\end{equation}
	where a non-zero $\nu_2$ indicates nontrivial AFHOT. This index serves the same role as symmetry indicators in static higher-order topological insulators/superconductors \cite{Ono2019,Skurativska2020}. We have therefore proved that all 2D $C_n$-symmetric AFHOTSC is ${\mathbb Z}_2$ classified in symmetry class D for an even $n$. Our complete topological classification is summarized in Table~\ref{Tab:classification}.

	\begin{figure}
	   \includegraphics[width=0.45\textwidth]{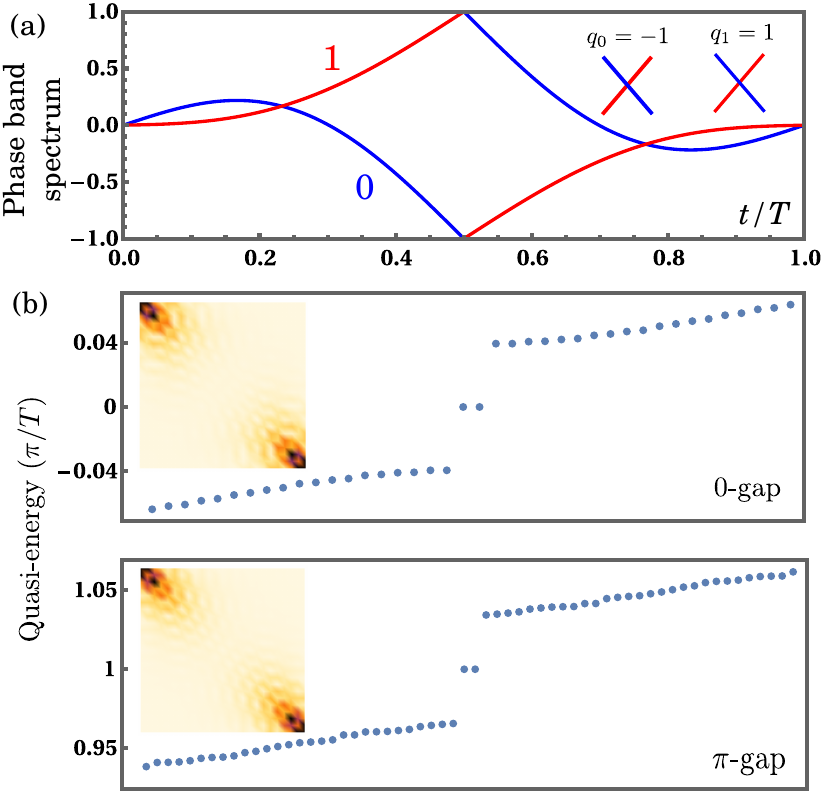}
	   \caption{(a) Phase bands at the $M$ momentum with $C_2$ eigenvalues 0 and 1. The phase band crossing is actually two overlapping Weyl points shown in the inset.  (b) Quasi-energy spectrum of the time-evolution unitary on a $25\times25$-site open-boundary square focusing on the 0 and $\pi$ gaps. The insets show the density distribution of the respective in-gap modes.}\label{fig:modelC2}
    \end{figure}

    \begin{figure}
	   \includegraphics[width=0.45\textwidth]{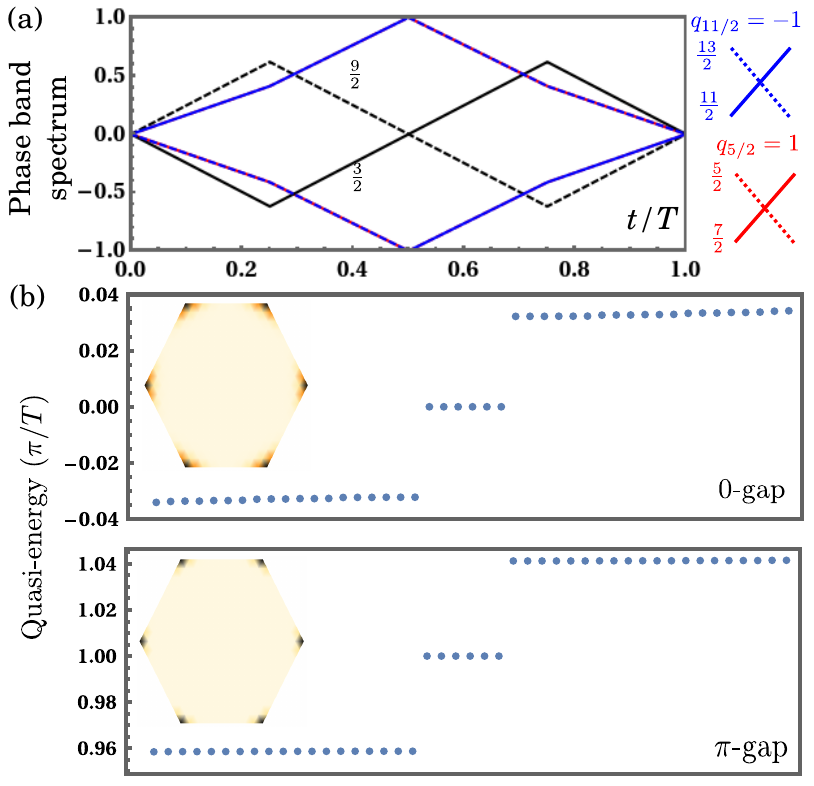}
	   \caption{(a) Phase bands at $\Gamma$ momentum with $C_6$ eigenvalues denoted. The inset on the right zooms in the $\pi-$ gap nodal points. (b) Quasi-energy spectrum of a time-evolution unitary simulated on a 271-site open-boundary hexagon. The insets show the density distribution of the respective in-gap modes.}\label{fig:modelC6}
   \end{figure}
	
	\textit{$C_2$-symmetric Floquet superconductor model - } To demonstrate our classification, we provide a minimal tight-binding model of $C_2$-symmetric AFHOTSC. The time-dependent Hamiltonian is 
	\begin{equation}
		\begin{split}
			&H(\mathbf{k},t) = \Gamma_{45}\left(4J_0-2J_0\cos k_x-2J_0\cos k_y -\mu \right) \\ &\quad+\Delta\left( \Gamma_{35}\sin k_x +\Gamma_5\sin k_y \right)
			+m\left( \Gamma_1+ \Gamma_{13}\right) \\
			& \quad -\Gamma_{45}  \left(4-2\cos k_x-2\cos k_y \right)J_D\cos\omega t .
		\end{split}
	\end{equation}   
	This model is equivalent to a pair of coupled first-order anomalous Floquet TSCs~\cite{Yang2018} with oppositely propagating Floquet chiral Majorana edge modes. The Hamiltonian has a PH symmetry $\mathcal{P}=\Gamma_4\mathcal{K}$ and $C_2=\Gamma_{45}$ [here $(C_2)^2=1$], which corresponds to $n=2$ and $\gamma=1$ ($J_1=0$, $J_2=1$). We denote $C_2$-invariant momenta $\bb{k}_0$ as $\Gamma=(0,0)$, $X=(\pi,0)$, $Y=(0,\pi)$ and $M=(\pi,\pi)$. We fix $\Delta=J_D/2=J_0$, $\mu=-2J_0$, $m=0.2J_0$ and $\omega=8J_0$. 
    For these parameters, we have $q^X_0=q^Y_0=q^M_0=1$, $q^X_1=q^Y_1=q^M_1=-1$ [see Fig.~\ref{fig:modelC2} (a) for the phase band spectrum at $M$] and thus $\nu_2=1$ bases on Eq.~\ref{eq:nu2}, indicating the presence of anomalous Floquet Majorana corner modes. This prediction is consistent with our numerical simulation on an open boundary geometry shown in Fig.~\ref{fig:modelC2} (b), where the quasienergy spectrum unambiguously shows in-gap modes at both the zero and $\pi$ gaps with density distributions localized at two $C_2$-related corners.
    
	\textit{$C_6$-symmetric Floquet superconductor model - } To further verify our classification scheme, we provide a Floquet tight-binding model with a piecewise driving Hamiltonian: $H(\mathbf{k},t) = \Delta_0 H_0$ for $t<T/4$ or $t>3T/4$, $H(\mathbf{k},t) = \Delta_1 H_1(\mathbf{k})$ for $T/4\le t \le 3T/4$, where
	\begin{equation}
      \begin{split}
		  & H_0 = \begin{pmatrix}
		   0 & i\sigma_0 & -\sigma_3 \\
		   -i\sigma_0 & 0 & i\sigma_0\\
		   -\sigma_3 & -i\sigma_0 & 0
		  \end{pmatrix},\\
		  & H_1(\mathbf{k}) = \underset{1\le i\le 3}{\oplus} \cos(\mathbf{k}\cdot\mathbf{a}_i)\sigma_3 + \cos(\mathbf{k}\cdot\mathbf{a}_i)\sigma_2.
		\end{split}
	\end{equation}  
	In the $(x,y)$-coordinate, we can define $\mathbf{a}_1=(1,0)$, $\mathbf{a}_2=(1/2,\sqrt{3}/2)$ and $\mathbf{a}_3=(-1/2,\sqrt{3}/2)$. The particle-hole and rotational symmetries are
	\begin{equation}
		\mathcal{P}= \begin{pmatrix}
		\sigma_1 & 0& 0 \\
		0 & \sigma_1 & 0\\
		0 & 0 & \sigma_1
		\end{pmatrix}\mathcal{K}, \quad C_6=\begin{pmatrix}
		0 & 0& -i\sigma_3 \\
		\sigma_0 & 0 & 0\\
		0 & \sigma_0 & 0
		\end{pmatrix},
	\end{equation}
	corresponding to the category of $(C_6)^6 = -1$ and $\gamma=0$, capable of hosting anomalous Floquet corner modes. The static second-order topology of the similar geometry has been studied in \cite{Benalcazar2014,Roberts2020} For the parameters $\Delta_0T=2\pi/\sqrt{2}$ and $\Delta_1T=-3\pi/\sqrt{2}$, the $\pi-$gap Weyl nodes only exist along the momentum $\Gamma=(0,0)$ and we find that $q_{5/2}=-q_{11/2}=1$ [see Fig.~\ref{fig:modelC6}(a)], indicating a six-fold set of corner Majorana modes at the zero and $\pi$ quasi-energy gaps. The simulation on an open-boundary hexagon, as shown in Figs.~\ref{fig:modelC6}(b) indeed confirms this prediction. We emphasize that in both examples of $C_2$ and $C_6$-models, the choice of parameters renders the Floquet phase bands trivial so the observed corner Majorana modes are inherently dynamical (see \cite{Supplement} for bulk-boundary correspondence in the general case).
	
	\textit{Conclusion - }We develop a general theory for classifying and characterizing superconductors with anomalous dynamical Floquet higher-order topology. By deciphering the hidden topological information encoded in the time-evolution phase, we establish phase-band singularities as a necessary signature for general AFHOTSCs. Through a phase-band dimensional reduction procedure, we construct a $\mathbb{Z}_2$ Floquet higher-order topological index, which exactly predicts the presence or absence of Floquet Majorana corner modes for a driven superconductor~\footnote{Note that our classification scheme has ignored the effect of translational symmetries. Thus, weak Floquet topological phases are not included in our classification in Table.~\ref{Tab:classification}.}.
	An interesting extension of this paper is to generalize the classification to BdG systems with other point group symmetries and nonsymmorphic spacetime symmetries~\cite{peng2019floquet,peng2020floquet}.
	
	\begin{acknowledgments} \textit {Acknowledgment - }This work is supported by the Laboratory for Physical Science. R.-X. Z. acknowledges the support from a JQI Postdoctoral Fellowship at the University of Maryland. Z.-C.Y. acknowledges funding by the NSF PFCQC program.
	\end{acknowledgments}
	
	\bibliographystyle{apsrev4-1}
	\bibliography{Floquet_SC}
\end{document}

\widetext{\section*{Supplemental material for ``Superconductors with anomalous Floquet higher-order topology"}}
		
		\section{Off-axis and non-linear degenerate points}
		%\section{Off-axis Weyl points}
		In this section, we study Weyl points located at $C_n$-related locations that are away from the rotational axis. We show that these singularities can be symmetrically moved to the rotational axis, after which they become either PH-invariant Weyl points or PH-related Weyl pairs.
			%mapped uniquely to on-axis singularities. 
%Once off-axis Weyl points have been mapped to on-axis ones, particle-hole rotation now either transforms the degenerate point within itself (PH-invariant Weyl point) or to a different point (PH-invariant Weyl pair). 

%We also show that PH-invariant degenerate points with non-linear dispersion can be decomposed into a collection of PH-invariant linear Weyl points and pairs. As a results our classification based on rotationally-invariant linear Weyl points essentially covers all possible cases of phase band singularities. 
%In the next two subsections studying off-axis nodal points, we temporarily neglect the particle-hole symmetry and only focus on the rotational symmetry.
		
		\subsection{Generic points}
		If a phase band degenerate point exists away from the rotational axis, there are necessarily another $n-1$ copies of this point related by the $C_n$-symmetry. The set thus has net topological charge $\pm n$.
%and can potentially alternate the topological properties of the Floquet system. 
The $C_n$-symmetric collection of Weyl points can be expressed as
		\begin{equation}
			h(k_x,k_y;t) = \pi\mathbb{I}_2 + v_t t \sigma_3 + (k_+^{n+u}-ck_+^u)\sigma_+ + (k_-^{n+u}-c^*k_-^u)\sigma_-, 
\label{eq:weyl}
		\end{equation}
		with $C_n = e^{iJ\theta_0}\text{diag}(1, e^{iu\theta_0})$, $\theta_0=\frac{2\pi}{n}$, $u>0$, $k_{\pm}=k_x\pm ik_y$ and $\sigma_\pm = \sigma_1\pm i\sigma_2$. It is easy to see that Hamiltonian~(\ref{eq:weyl}) has a Weyl point at $k=0$ with charge $-{\rm sgn}(v_t) u$, and $n$ additional off-axis Weyl points at $(k,\theta_k)=(|c|^{1/n},\text{arg}(c)/n+2\pi l /n)$ with $l\in \{0,\dots,n-1\}$.
%beside an on-axis singularity for $u>0$, there exist $n$ off-axis Weyl points at $(k,\theta_k)=(|c|^{-1},\text{arg}(c)/n+2\pi l /n)$ with $l\in \{0,\dots,n-1\}$.
By continuously decreasing $c\to 0$, %no singularities are created or destroyed but 
the off-axis Weyl points are continuously moved to the rotational axis, resulting in a rotationally invariant degenerate point with a dispersion of order $n+u$. This high-order-dispersion Weyl point has topological charge $-\text{sgn}(v_t)(n+u)$, exactly the same as the total charge of the initial configuration. In the third part of this section, we show that a high-order dispersion nodal point can be further decomposed into a set of PH-invariant linear nodal points and pairs.
		
	\subsection{High-symmetry points}
\begin{figure}
			\centering
			\includegraphics[scale=0.4]{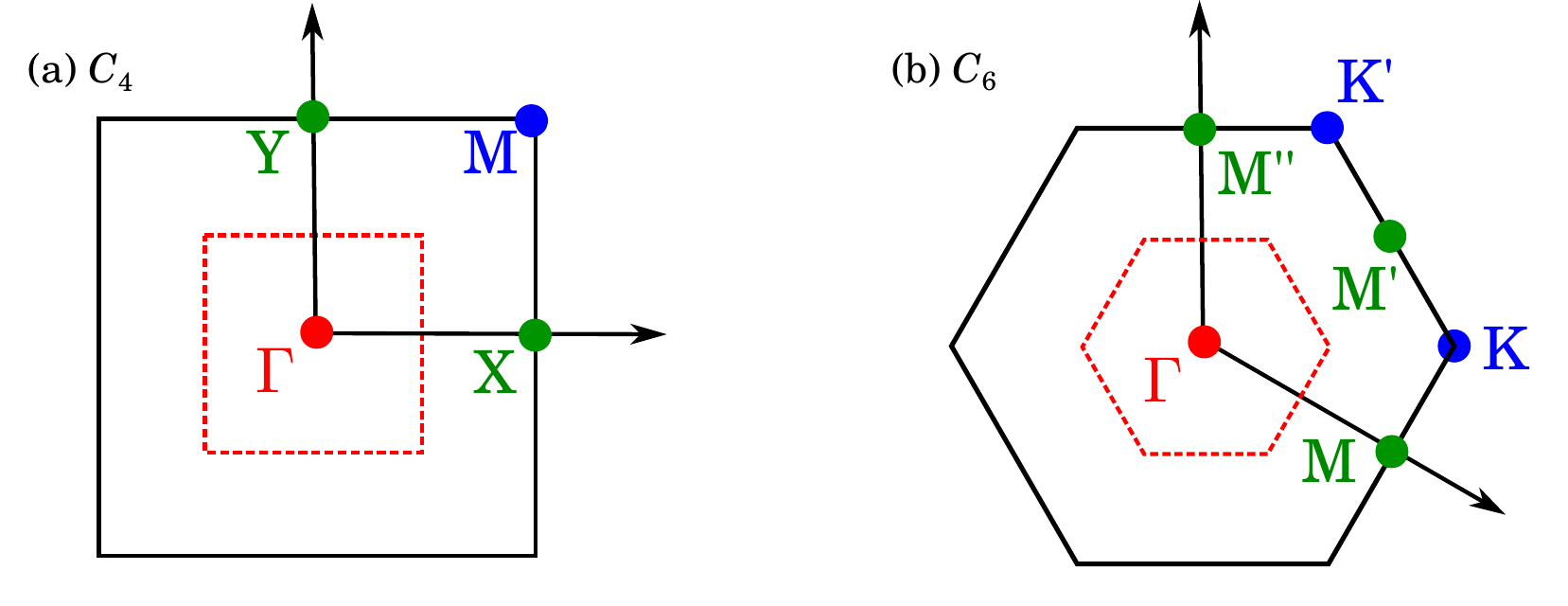}
			\caption{Brillouin zone and high-symmetry momenta for (a) 4-fold and (b) 6-fold rotations. The red dashed line represents the BZ folded by half, corresponding to the doubled unit cell. In that case, the original high-symmetry momenta can be identified with $\Gamma$.\label{fig1}}
		\end{figure}

The above argument cannot be directly applied to singularities that exist on high-symmetry but non-$C_n$ symmetric points in the Brillouin zone (BZ). Examples include the $X$ and $Y$ points of the square BZ, which are $C_2$ invariant; the $K$ and $K'$ points of the hexagonal BZ, which are $C_3$ invariant; the $M$, $M'$, and $M''$ points of the hexagonal BZ, which are $C_2$ invariant (see Fig.~\ref{fig1}). Weyl points located at these high-symmetry momenta carry eigenvalues of the corresponding little group at these points. As we pointed out in the main text, we focus on strong topological phases that exist even when translational symmetries are broken. Therefore, these high-symmetry points can always be folded to the $\Gamma$ point when the system is not translational invariant. Below we demonstrate explicitly how these Weyl points can be folded to the $\Gamma$ point when the unit cell of the original system doubles or triples, using the hexagonal BZ as an example. The square BZ can be analyzed in a similar manner.\\

	\textbf{(a) $K$ and $K'$}
		
   For $C_6$ symmetry, $K$ and $K'$ are related by a $C_6$ rotation and invariant under a $\mathbb{Z}_3$ subgroup $\{\mathbf{1},C_6^2, C_6^4 \}$.  
   If we express the collective Weyl points at $K$ and $K'$ as $h_K\oplus h_{K'}$ with $h_\mathbf{A}=v_t t \sigma_3 + (k_x - A_x)\sigma_1-(k_y-A_y)\sigma_2$, then this singularity can be labeled by $J\in\{0,1,2\}$, corresponding to
	    \begin{equation}
	    	C_6^2 = \tau_0\otimes \left[ e^{i2\pi J/3}\text{diag}\left(1, e^{i2\pi/3}\right)  \right],
	    \end{equation}
		with $\tau_{0,1,2,3}$ being the Pauli matrices acting on the valley degree of freedom. Accordingly, we derive
		\begin{equation}\label{eq:C4}
			C_6 = \tau_1 \otimes \left[\pm e^{i\pi J/3}\text{diag}\left(1, e^{i\pi/3}\right)\right]. 
		\end{equation}
 After tripling the unit cell (the BZ is reduced to a third), $K,K' \to \Gamma$ and the singularity is now invariant under $C_6$. We note that the ambiguity $\pm 1$ in Eq.~\eqref{eq:C4} does not alter the eigenvalue spectrum as $\sigma_1$ also has eigenvalues of $\pm 1$. It is straightforward to rewrite the singularity in the basis of $C_6$ eigenvectors (by diagonalizing Eq.~\eqref{eq:C4}); as a result, we obtain two $C_6$-symmetric Weyl points at $\Gamma$ in the basis $\left(\psi_J,\psi_{J+1} \right)^T$ and $\left(\psi_{J+3},\psi_{J+4} \right)^T$\\
		
		\textbf{(b) $M, M'$ and $M''$}\\
		Unlike $K$ and $K'$, the $M, M'$ and $M''$ points are invariant under a $\mathbb{Z}_2$ subgroup $\{\mathbf{1},C_6^3\}$. The collective singularity $h_M\oplus h_{M'}\oplus h_{M''}$ can be labeled with $J\in\{0,1\}$, corresponding to
		\begin{equation}
			C_6^3 = \mathbf{I}_3\otimes\left(e^{i\pi J}\sigma_3 \right), \quad \text{and } C_6 = \begin{pmatrix}
			0 & 0 & 1\\ 1& 0 & 0 \\ 0 & 1& 0
			\end{pmatrix}\otimes\left[\omega e^{i\pi J/3} \text{diag}\left(1, e^{i\pi/3}\right)\right],
		\end{equation}
		where $\omega$ is one of the three solutions of $\omega^3=1$. Again, this ambiguity does not alter the eigenvalue spectrum. After folding the BZ and rewrite the degenerate point in $C_6$ eigenvectors, we obtain three $C_6$-symmetric Weyl points in the basis $\left(\psi_J,\psi_{J+1} \right)^T$, $\left(\psi_{J+2},\psi_{J+3} \right)^T$ and $\left(\psi_{J+4},\psi_{J+5} \right)^T$.
		
		\subsection{Non-linear dispersion singularities}
		
		We now show that a non-linear singularity is adiabatically connected to a combination of linear nodal points and pairs by adding trivial bands. It is, in fact, easier to construct the argument in the opposite direction, i.e. a suitable collection of linear nodal points and pairs can be transformed into a non-linear nodal point plus gapped trivial bands. We consider a nodal point in the basis $\left(\psi_{J},\psi_{J+1}\right)^T$ where $\bar{J} = J+1$
		\begin{equation}
			h_1=v_tt\sigma_z+v_k(k_x\sigma_x-k_y\sigma_y);
		\end{equation}
		and an additional nodal pair in the basis $(\psi'_{J-1},\psi'_{J},\psi'_{J+1},\psi'_{J+2})^T$
		\begin{equation}
			\begin{split}
				h_2&=\tau_0\otimes\left[v_tt\sigma_z+v_k(k_x\sigma_x-k_y\sigma_y)\right]
			\end{split}
		\end{equation}
		Here $\overline{J'-1}=\bar{J'}+1\equiv J'+2$, thus satisfy the PH symmetry. The combined Hamiltonian $h=h_1\oplus h_2$ has $C_n=\exp\left[i\theta_0\text{diag}(J,J+1,J-1,J,J+1,J+2)\right]$ and the particle-hole symmetry
		\begin{equation}
			\mathcal{P}=\mathcal{K}\begin{pmatrix}
				\sigma_x & 0 & 0\\
				0 & 0 & \sigma_x\\
				0 & \sigma_x & 0
			\end{pmatrix}
		\end{equation}
		Then, it is permissible to add a mass term
		\begin{equation}
			h_m = m\begin{pmatrix}
				0 & 0 & 0 & 1 &  0  & 0\\
				0 & 0 & 0 & 0 & -1  & 0\\
				0 & 0 & 0 & 0 &  0  & 0\\
				1 & 0 & 0 & 0 &  0  & 0\\
				0 & -1 & 0 & 0 & 0 & 0\\
				0 & 0 &0 &0 &0 &0
			\end{pmatrix},
		\end{equation}
		where it is straightforward to check that $[C_n,h_m]=0$ and $\{\mathcal{P},h_m\}=0$. For $k_x=k_y=t=0$, the degenerate subspace reduces to $\left(\psi'_{J-1},\psi'_{J+2}\right)^T$. We can construct the effective low-energy Hamiltonian at the vicinity of this point as follows. The diagonal term can be obtained from the first-order perturbation
		\begin{equation}
			\begin{split}
				&\bra{\psi'_{J-1}}h_{eff}\ket{\psi'_{J-1}} \approx \bra{\psi'_{J-1}}h\ket{\psi'_{J-1}} = v_tt\\
				&\bra{\psi'_{J+2}}h_{eff}\ket{\psi'_{J+2}} \approx \bra{\psi'_{J+2}}h\ket{\psi'_{J+2}} = -v_tt.
			\end{split}
		\end{equation} 
		The off-diagonal term, however, requires the third-order perturbation for non-zero contribution
		\begin{equation}
			\begin{split}
				&\bra{\psi'_{J-1}}h_{eff}\ket{\psi'_{J+2}}\approx \sum_{\alpha,\beta} \frac{\bra{\psi'_{J-1}}h\ket{\alpha}\bra{\alpha}h\ket{\beta}\bra{\beta}h\ket{\psi'_{J+2}} }{E_\alpha E_{\beta}}=\frac{k_+^3}{m^2},
			\end{split}
		\end{equation}
		where $\alpha=(\psi_{J}\pm\psi'_{J})/\sqrt{2}$ corresponding to $E_\alpha=\pm m$, and $\beta=(\psi_{{J+1}}\pm\psi'_{{J+1}})/\sqrt{2}$ corresponding to $E_\beta=\mp m$. The effective Hamiltonian exactly represents a cubic nodal point. The result indeed agrees with the conservation of topological charge. This suggests that a non-linear nodal point of charge $q$ in the subspace $\left(\psi_{J},\psi_{J+|q|}\right)^T$ can be decomposed into one linear nodal point and $m=(|q|-1)/2$ linear nodal pairs. Schematically,
		\begin{equation}
			\begin{split}
				& X^{\pm q}_{J,J+|q|} \sim X^{\pm 1}_{J+m,\bar{J}-m}\oplus\sum_{l=0}^{m-1}X^{\pm 2}_{J+l,J+l+1,\bar J-l-1,\bar J-l},
			\end{split}
		\end{equation}
		where $X^q_{\{J\}}$ denotes a singularity of the $C_n$-invariant subspace $\{J\}$ carrying monopole charge $q$. Here $m$ is always an integer since $|q|$ must be odd and if $\bar J\equiv J+|q|$ then $\bar{J}-m\equiv J+m+1$ satisfying the rotational and PH symmetries. For a nodal pair with non-linear dispersion, a similar decomposition rule based on the conservation of topological charge can be made with the only difference is now the PHS maps one block matrix to another
		\begin{equation}
			\begin{split}
				&X^{\pm 2q}_{J,J+|q|,\bar J -|q|,\bar J} \sim \sum_{l=0}^{|q|-1}X^{\pm 2}_{J+l,J+l+1,\bar J-l-1,\bar J-l}.
			\end{split}
		\end{equation}
		Additionally if the sequence of $l$ wraps several times around $n$, at $l$ such that $J+l\equiv \bar J-l$ mod $n$, the PH nodal pair can be further decomposed into two nodal point with a closed PH mapping
		\begin{equation}
			X^{\pm 2}_{J+l,J+l+1,\bar J-l-1,\bar J-l} \sim X^{\pm 1}_{J+l,\bar J -l}\oplus X^{\pm 1}_{\bar J -l-1,J+l+1}.
		\end{equation}
		
		In conclusion, all the topological singularity in a 2D Floquet class D can be described by linear nodal points in the subspace $\left(\psi_{J},\psi_{J+1}\right)^T$ with $\bar{J} \equiv J+1$ or linear nodal pairs belonging to subspace $\left(\psi_J,\psi_{J+1},\psi_{\bar J -1}, \psi_{\bar J}\right)^T$ with $\bar{J} \not\equiv J+1$.
		
		\section{Dimensional reduction}
		In this section, we present the details for the phase band dimensional reduction when the first-order topology is trivial, i.e. the net topological charge is zero. We consider two elementary types of composite zero-net-charge singularities: (i) two Weyl points of opposite charges formed by 4 phase bands, and (ii) two Weyl pairs or one Weyl pair and two Weyl points formed by 8 phase bands. The general case is simply a combination of these elementary cases, and the total number $d$ of phase bands involved is either $8n$ or $8n+4$. 
		
	   \subsection{$d=4$}
		We first demonstrate the dimensional reduction for a zero-net-charge singularity formed by two Weyl points of opposite charge
	    \begin{equation}\label{eq:2Weylpoints}
		\tilde{h}(k_x,k_y;t) = \pi\mathbb{I}_4 + v_k(k_x\Gamma_4-k_y\Gamma_5)+ v_t t \Gamma_{3}.
		\end{equation} 
		A constant PH-invariant mass term $\Gamma_1$ is not allowed by the rotational symmetry. However, we can modulate the mass term in the manner defined in the main text so that $C_n m(r,\theta) C_n^{-1} = -m(r,\theta)=m(r,\theta+\theta_0)$. The domain walls at $\theta = l\theta_0$ are gapless. Without loss of generality, we study the domain wall at $\theta=0$, which is parallel to the $x-$axis. The wavefunction $\psi$ of the in-gap domain wall mode can be solved by rewriting Hamiltonian \eqref{eq:2Weylpoints} with $k_y\to -i\partial_y$ and setting $k_x=t=0$ at the degenerate point. This directly leads to a zero-mode equation
		\begin{equation}
		\left[v_k\partial_y + m_0\Gamma_{15}\text{sgn}(y) \right] \psi = 0.
		\end{equation}
		Using the ansatz $\psi=e^{f(y,\eta)}\psi_{\eta}$ where $\Gamma_{15}\psi_{\eta}=\eta\psi_{\eta}$ and $\eta=\pm 1$, one can obtain that $f(y,\eta)=-m_0\eta |y|/v_k.$ 
		Thus, for a normalizable wavefunction, we require that $\eta=\text{sgn}(m_0/v_k)$, constraining the zero-mode subspace to always be two-dimensional ($\Gamma_{15}$ has two positive and two negative eigenvalues). It is then straightforward to arrive at an effective domain-wall Hamiltonian
		$\tilde{h}_{1D}(k_x,t) = \pi\mathbb{I}_2 + v_kk_x\sigma_1 + v_t t \sigma_3$, the gaplessness of which is protected by the PH symmetry $\mathcal{P}_{1D}=\mathcal{K}\sigma_1$ with ${\cal K}$ being the complex conjugation operation. \\
		
		\subsection{$d=8$}
		We now move to a subspace containing one nodal pair and another pair or two points of opposite topological charges. In the basis $\left(\psi_{J_b},\psi_{J_b+1}, \psi_{J_a},\psi_{J_a+1},\psi_{\bar J_b}, \psi_{\bar{J}_b-1},\psi_{\bar J_a},\psi_{\bar J_a -1}\right)^T$, the effective Hamiltonian takes the form
		\begin{equation}
			\tilde{h}(k_x,k_y;t)=\pi\mathbb{I}_8 + \left[v_t t \Gamma_{3} + v_k(k_x\Gamma_4-k_y\Gamma_5) \right]\oplus \left[-v_t t \Gamma_{3} +v_k(k_x\Gamma_4+k_y\Gamma_5) \right].
		\end{equation}
		A general PH-invariant mass term that can open a gap is
		\begin{equation}
			m=\begin{pmatrix}
				0 & c \sigma_3\\
				c^* \sigma_3& 0
			\end{pmatrix} \oplus \begin{pmatrix}
				0 & -c^* \sigma_3\\
				-c \sigma_3 & 0
			\end{pmatrix}
		\end{equation} 
		where $c$ is complex number. This mass term, however, is not allowed by the rotational symmetry when $J_a \neq J_b$. In fact
		\begin{equation}
			C_n m C_n^{-1} =\begin{pmatrix}
				0 & c e^{i\Delta\theta}  \sigma_3\\
				c^* e^{-i\Delta\theta}  \sigma_3& 0
			\end{pmatrix} \oplus \begin{pmatrix}
				0 & -c^*e^{-i\Delta\theta} \sigma_3\\
				-ce^{i\Delta\theta} \sigma_3 & 0
			\end{pmatrix} \neq m,
		\end{equation} 
		where $\Delta\theta=2\pi(J_b-J_a)/n$.
		To recover global symmetry, one can make the complex factor $c$ spatially modulated
		\begin{equation}
			c(r,\theta)=c_0e^{-i\Delta\theta\lfloor \theta/\theta_0 \rfloor}.
		\end{equation}
		The dimensional reduction process can be performed simultaneously in a similar fashion for the ``electron" and ``hole" blocks, so we only consider here the former. Upon applying the mass term, the phase band is gapped in bulk away from $\theta=l\theta_0$ with $l$ being an integer. At a domain wall, for example $\theta=0$ (perpendicular to the $y-$axis), one can substitute $k_y\to -i\partial_y$ and set $t,k_x=0$ to study the zero-energy modes through the equations
		\begin{equation}
			\begin{split}
				& y>0: \quad \left[v_k\partial_y + c_0\Gamma_{15}\right]\ket{\psi} =0 \\
				&  y\le 0: \quad\left[v_k\partial_y + c_0\left(\Gamma_{15}\cos\Delta\theta - \Gamma_{25}\sin\Delta\theta\right)\right]\ket{\psi} =0 
			\end{split}.
		\end{equation}
		Assuming $c_0>0$, we obtain the piece-wise normalizable solutions as follows
		\begin{equation}
			\begin{split}
				&  y>0: \quad \ket{\psi} = e^{-c_0y/v_k}\left[a_1\left(1,0,0,1\right)^T + a_2\left(0,1,1,0\right)^T \right]\\
				&  y\le 0: \quad  \ket{\psi} = e^{c_0y/v_k}\left[b_1\left(-e^{i\Delta\theta},0,0,1\right)^T + b_2\left(0,-e^{i\Delta\theta},1,0\right)^T \right]
			\end{split}.
		\end{equation}
		For $J_b-J_a\neq n/2$ mod $n$, the two pieces at the boundary cannot be joined continuously, implying that the domain wall must be gapped. On the other hand, if $J_b-J_a= n/2$, a pair of zero-energy solution can be obtained by setting $a_1=b_1$ and $a_2=b_2$, leading to the effective (1+1)D Dirac Hamiltonian
		\begin{equation}
			\tilde{h}_{1D}(k_x,t) = \pi\mathbb{I}_4 + (v_t t\sigma_3+v_k k_x\sigma_1)\otimes(-v_t t\sigma_3+v_k k_x\sigma_1) = \pi\mathbb{I}_4+v_t t \Gamma_3+v_k k_x \Gamma_{4},
		\end{equation}
		with the PH symmetry $\mathcal{P}_{1D}=\Gamma_{23}\mathcal{K}$. Even though the domain wall is seemingly gapless, the gap can be opened by a perturbation $\sim\Gamma_{2,34}$. Therefore, for any $J_a\ne J_b$, the domain wall is always gapped, leading to trivial boundary behavior. 
		
		\section{General bulk-boundary correspondence}
			
	   The model Hamiltonian we presented in the main text has trivial Floquet bulk bands, so the corner modes are entirely generated by the non-trivial anomalous Floquet topology. In the general case, the exact numbers of $0$ and $\pi$ Floquet Majorana modes, denoted as ${\cal N}_0$ and ${\cal N}_{\pi}$, are actually predicted by the combination of our anomalous Floquet topological index $\nu_2$ and a static topological index $\nu_s$ as
	      \begin{equation}\label{eq:corner index}
	         \mathcal{N}_{\pi}=\nu_2, \quad \mathcal{N}_{0}=\nu_2+\nu_s\ \  \md 2.
	      \end{equation}
	  Here, $\nu_s$ is the $C_n$-symmetry index defined for the bulk Floquet bands and related to boundary modes at the zero gap.
	
		\subsection{$C_2$-symmetric Floquet superconductor}
\begin{figure}
		 \includegraphics[scale=0.5]{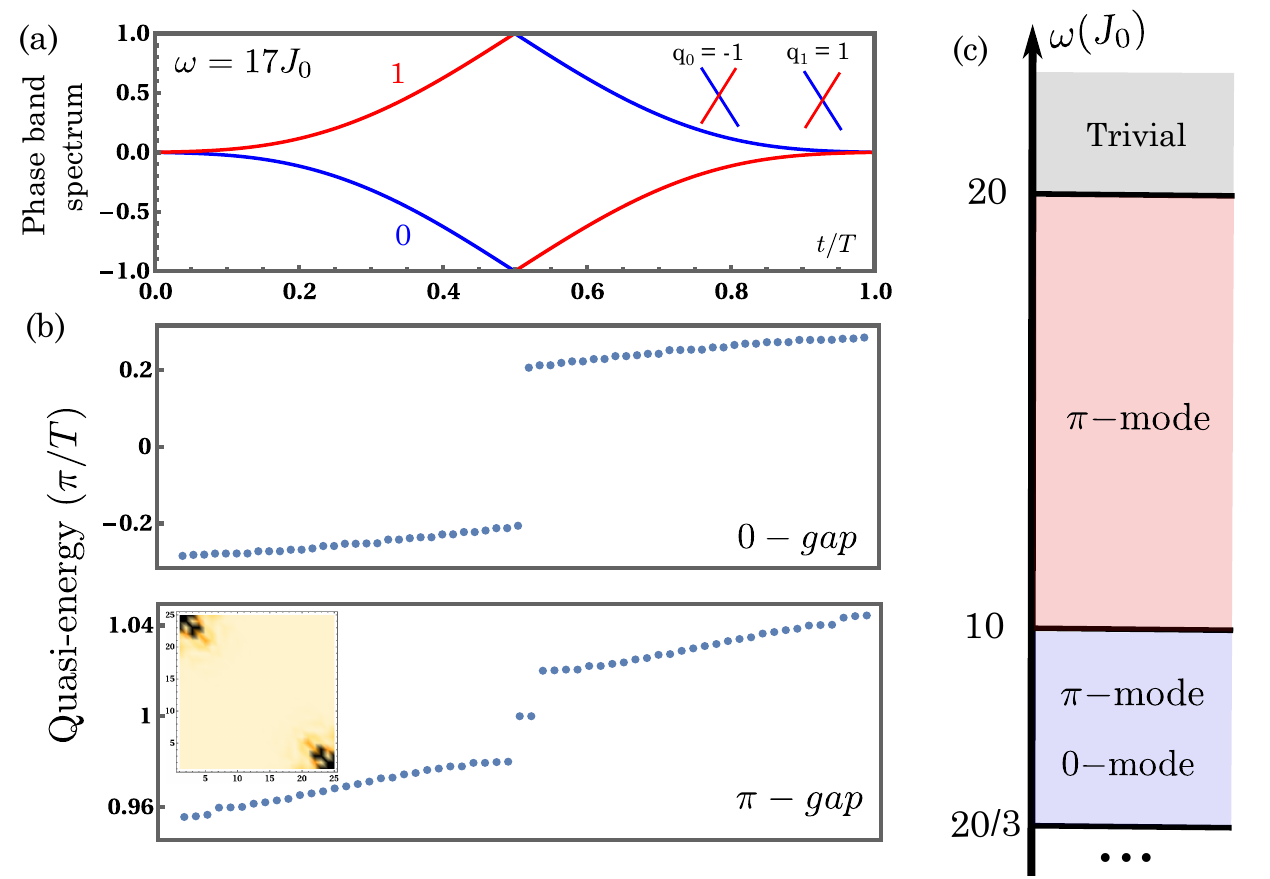}
		 \caption{(a) Phase band spectrum at momentum $M$ where the crossing is two overlapping Weyl points of opposite charges as shown in the inset. (b) Quasi-energy spectrum of the time evolution operator on an open-boundary geometry. The 0-gap lacks boundary modes while the $\pi$-gap carries two $C_2$-related localized corner modes. The inset shows the density profile of the boundary modes. (c) Phase diagram of the boundary property as $\omega$ decreases.}\label{fig:model}
		\end{figure}

        As an example of non-trivial Floquet bands, we reuse the model Hamiltonian introduced in the main text with the only change being $\omega=17J_0$.
		With this value of driving frequency, $q^M_0 = 1$ and $q^M_1=-1$ [See Fig.~\ref{fig:model}(a) for the phase band at $M$ momentum]; thus $\nu_2=1$ indicating anomalous higher-order Floquet topology. 
		
        On the other hand, the properties of the Floquet bands can be conveniently studied similarly to a static superconductor from the symmetry data, which is the $C_n$-eigenvalues of the occupied (below zero energy) bands at high-symmetry momentum. At the momenta $\Gamma, X, Y, M$, the symmetry data for $\omega=17J_0$ and $\omega=8J_0$ are $(--,--,--,++)$ and $(--,++,++,--)$, respectively. As in Ref.~\cite{Skurativska2020}, the static topological invariant $\nu_s$ can be obtained as
	\begin{equation}
		\begin{split}
			&\nu_s = \sum_{\mathbf{k\in \text{TRIM}}} \frac{n_{\mathbf{k}}^+ - n_{\mathbf{k}}^-}{2}\text{ mod } 2,
		\end{split}
	\end{equation}
    with $n_{\mathbf{k}}^\alpha$ being the number of occupied bands with eigenvalue $\alpha$ at the time-reversal invariant momentum $\mathbf{k}$.  
	For $\omega = 17$, $\nu_2 = n_s = 1$, thus we expect no $0-$gap modes and one pair of $\pi-$gap modes, which agrees with numerical simulation in Fig.~\ref{fig:model}(b). On the other hand, the case $\omega=8J_0$ presented in the main text has $\nu_s=0$, showing that Floquet bands are incapable of generating the observed higher-order topology. Combining the phase band index $\nu_2$ and the Floquet band index $\nu_s$, we give the phase diagram of first few topological phases with the corresponding boundary signatures as $\omega$ decreases in Fig.~\ref{fig:model}(c).
		
	\subsection{$C_6$-symmetric Floquet superconductor}	
\begin{figure}
	\includegraphics[scale=0.7]{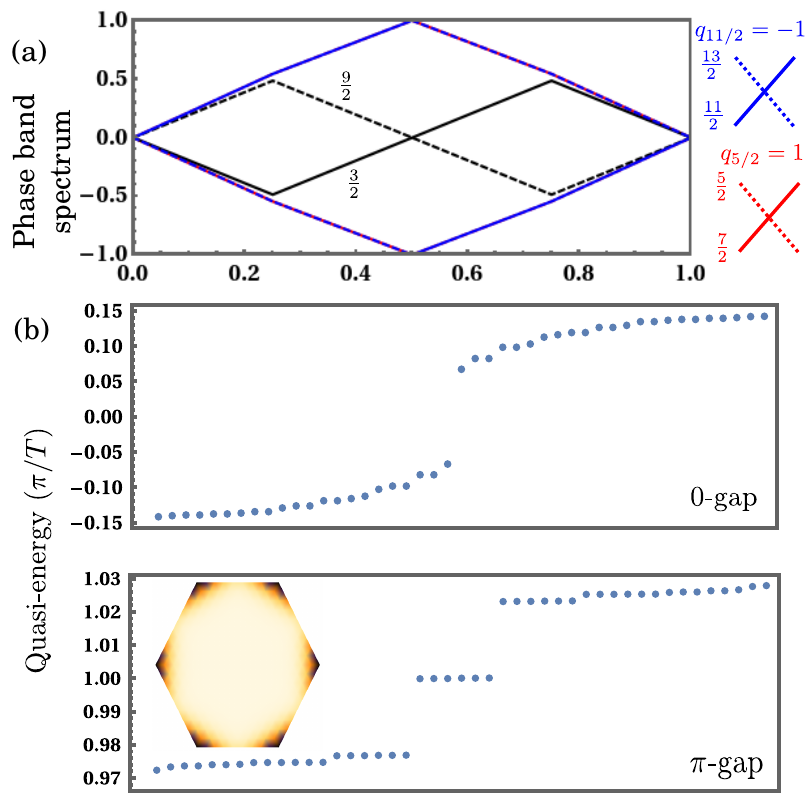}
	\caption{(a) Phase band spectrum at momentum $\Gamma$ where the crossing is two overlapping Weyl points of opposite charge. (b) Quasi-energy spectrum of the time evolution operator on an open-boundary hexagon showing the absence(presence) of the boundary modes at the 0($\pi$) gap.}\label{fig:modelC6}
   \end{figure}

   The same Floquet analysis can be performed for the $C_6$-models demonstrated in the main text. By choosing $\Delta_0T=2\pi/\sqrt{2}$ and $\Delta_1T=-3\pi/2\sqrt{2}$, we can introduce the non-trivial topology to the Floquet bands. The return map as in Fig.~\ref{fig:modelC6} shows that $q_{5/2}=-q_{11/2}=1$, indicating $\nu_2=1$. On the other hand, the topology of the $C_6$ Floquet bands can be analyzed through the symmetry data of occupied bands (quasi-energy $<0$) at high symmetry momenta. The momentum $\Gamma=(0,0)$ has six-fold symmetry and thus each band can be labeled by $n_6=1,\dots,6$ corresponding to the eigenvalue $e^{2i\pi (n+0.5)/6}$) with respect to $C_6$; meanwhile $M=(\pi,-\pi/\sqrt{3})$ has two-fold symmetry and thus the state labels are reduced to $n_2= 1,2$ corresponding to the eigenvalue $e^{i\pi (n_2+0.5)}$ with respect to $C_6^3$; and $K=(4\pi/3,0)$ has a three-fold symmetry, allowing a set of label $n_3=1,2,3$ corresponding to the eigenvalues $e^{2i\pi/3(n_3+0.5)}$ of $C_6^2$. We summarize the symmetry data at $\Gamma,K,M$ for the cases of $\Delta_1T=-3\pi/\sqrt{2}$ (trivial Floquet bands) and $\Delta_1T=-3\pi/2\sqrt{2}$ (non-trivial Floquet bands) as $(1,3,5;1,1,1;1,2,3)$ and $(1,3,5;1,2,2;1,2,3)$. According to the bulk-boundary correspondence derived in \cite{Roberts2020}
	\begin{equation}
		\nu_s = \frac{n_M^1 - (n_\Gamma^1+n_\Gamma^3+n_\Gamma^5)}{2}  \text{ mod } 2, 
	\end{equation}
	where $n_\mathbf{k}^i$ is the number of $i-$labeled occupied state at the momentum $\mathbf{k}$. By substituting the symmetry data into the static bulk-boundary correspondence, we find that for $\Delta_1T=-3\pi/\sqrt{2}$, $\nu_s=0$; while for $\Delta_1T=-3\pi/2\sqrt{2}$, $\nu_s=1$. Together with the anomalous Floquet index $\nu_2=1$, we can predict that the former hosts corner modes at both zero and $\pi$ gaps, consistent with simulation results in the main text; while the latter should only have corner modes at the $\pi$ gap, which is verified by numerical results shown in Fig.~\ref{fig:modelC6}(b).
		
	\bibliographystyle{apsrev4-1}
     \bibliography{Floquet_SC}